\documentclass[
reprint,
superscriptaddress,
frontmatterverbose, 
showpacs,
preprintnumbers,
nofootinbib,
amsmath,
amssymb,
prl,
floatfix,
twocolumn
]{revtex4-2}

\usepackage{soul}
\usepackage[inline]{enumitem}
\usepackage[utf8]{inputenc}
\usepackage[normalem]{ulem}
\usepackage{graphicx}
\usepackage{dcolumn}
\usepackage{bm}
\usepackage{color}
\usepackage[dvipsnames]{xcolor}
\usepackage[
    colorlinks = true,
    linkcolor = BlueViolet,
    anchorcolor = purple,
    citecolor = purple,
    filecolor = purple,
    urlcolor = BlueViolet]{hyperref}
    
\usepackage{url}
\usepackage{xspace}
\usepackage{slashed}
\usepackage{multirow,bigstrut}
\usepackage{mathrsfs} 
\usepackage{relsize,amsmath}
\usepackage[geometry]{ifsym}
\usepackage{amssymb}
\usepackage{pifont}
\usepackage{physics}

\usepackage{cleveref}

\usepackage{rotating}
\usepackage{ragged2e}

\usepackage{fontawesome} 
\definecolor{blue-violet}{rgb}{0.33, 0.17, 0.89}

\newcommand{\nano}{NANOGrav }

\newcommand{\titleprl}[1]{\vspace{0.3cm} \noindent \textit{\textbf{#1}}:}

\newcounter{CommentCount}
\definecolor{MH}{rgb}{0.0,0.6,9}
\definecolor{palatinate}{rgb}{0.494, 0.192, 0.482}

\definecolor{teal}{HTML}{008080}

\usepackage{siunitx}

\DeclareSIUnit \s {\second}
\DeclareSIUnit \ns {\nano\second}
\DeclareSIUnit \mus {\micro\second}
\DeclareSIUnit \ms {\milli\second}
\DeclareSIUnit \MB {\mega\byte}
\DeclareSIUnit \GB {\giga\byte}
\DeclareSIUnit \TB {\tera\byte}
\DeclareSIUnit \PB {\peta\byte}
\DeclareSIUnit \Mbps {\mega\bit/\s}
\DeclareSIUnit \Gbps {\giga\bit/\s}
\DeclareSIUnit \Tbps {\tera\bit/\s}
\DeclareSIUnit \Pbps {\peta\bit/\s}
\DeclareSIUnit \kton {\kilo\tonne} 
\DeclareSIUnit \kt {\kilo\tonne}
\DeclareSIUnit \Mt {\mega\tonne}
\DeclareSIUnit \eV {\electronvolt}
\DeclareSIUnit \keV {\kilo\electronvolt}
\DeclareSIUnit \MeV {\mega\electronvolt}
\DeclareSIUnit \GeV {\giga\electronvolt}
\DeclareSIUnit \TeV {\tera\electronvolt}
\DeclareSIUnit \PeV {\peta\electronvolt}
\DeclareSIUnit \EeV {\exa\electronvolt}
\DeclareSIUnit \m {\meter}
\DeclareSIUnit \cm {\centi\meter}
\DeclareSIUnit \in {\inchcommand}
\DeclareSIUnit \km {\kilo\meter}
\DeclareSIUnit \kV {\kilo\volt}
\DeclareSIUnit \kW {\kilo\watt}
\DeclareSIUnit \MW {\mega\watt}
\DeclareSIUnit \MHz {\mega\hertz}
\DeclareSIUnit \mrad {\milli\radian}
\DeclareSIUnit \year {years}
\DeclareSIUnit \POT {POT}
\DeclareSIUnit \sig {$\sigma$}
\DeclareSIUnit\parsec{pc}
\DeclareSIUnit\lightyear{ly}
\DeclareSIUnit\foot{ft}
\DeclareSIUnit\ft{ft}
\DeclareSIUnit \ppb{ppb}
\DeclareSIUnit \ppt{ppt}
\DeclareSIUnit \samples{S}
\DeclareSIUnit \pe{PE}
\DeclareSIUnit \T{T}

\newcommand{\enu}{\E_\enu}

\begin{document}

\title{Supercooled Dark Scalar Phase Transitions explanation of NANOGrav data}

\author{Francesco Costa}
\email{francesco.costa@matfyz.cuni.cz}
\affiliation{Institute of Particle and Nuclear Physics, Faculty of Mathematics and Physics,
Charles University in Prague, V Hole\v{s}ovi\v{c}k\'ach 2, 180 00 Praha 8, Czech Republic}

\author{Jaime Hoefken Zink}
\email{jaime.hoefkenzink@ncbj.gov.pl}
\affiliation{National Centre for Nuclear Research, Pasteura 7, Warsaw, PL-02-093, Poland}

\author{Michele Lucente}
\email{michele.lucente@unibo.it}
\altaffiliation[\\Also visitor at: ]{\emph{Theoretical Physics Department, Fermi National Accelerator Laboratory, Batavia, Illinois 60510, USA}}
\affiliation{Dipartimento di Fisica e Astronomia, Universit\`a di Bologna, via Irnerio 46, 40126 Bologna, Italy}
\affiliation{INFN, Sezione di Bologna, viale Berti Pichat 6/2, 40127 Bologna, Italy}

\author{Silvia Pascoli}
\email{silvia.pascoli@unibo.it}
\affiliation{Dipartimento di Fisica e Astronomia, Universit\`a di Bologna, via Irnerio 46, 40126 Bologna, Italy}
\affiliation{INFN, Sezione di Bologna, viale Berti Pichat 6/2, 40127 Bologna, Italy}

\author{Salvador Rosauro-Alcaraz}
\email{rosauroa@bo.infn.it}
\affiliation{INFN, Sezione di Bologna, viale Berti Pichat 6/2, 40127 Bologna, Italy}

\date{\today}

\begin{abstract}
The evidence of a Stochastic Gravitational Wave Background (SGWB) in the nHz frequency range is posed to open a new window on the Universe. A preferred explanation relies on a supercooled first order phase transition at the 100~MeV--GeV scale. In this article, we address its feasibility going from the particle physics model to the production of the gravitational waves. We take a minimal approach for the dark sector model introducing the fewest ingredient required, namely a new $U(1)$ gauge group and a dark scalar that dynamically breaks the symmetry. Supercooling poses challenges in the analysis that put under question the feasibility of this explanation: we address them, going beyond previous studies by carefully considering the effects of a vacuum domination phase and explicitly tracking the phase transition from its onset to its completion. We find that the proposed model can successfully give origin to the observed PTA SGWB signal. The strong supercooling imposes a correlation between the new gauge coupling and the scalar quartic one, leading to a significant hierarchy between the (heavier) gauge boson and the dark scalar. Ultimately, information on phase transitions from SGWB observations could provide a direct probe of the microphysics of the Early Universe and be used to guide future searches of dark sectors in laboratories.
\end{abstract}

\maketitle


\titleprl{Introduction} Gravitational waves (GWs) have opened an unprecedented observational window into the Universe. The recent birth of GW astronomy, by detecting GWs from binary black hole mergers by LIGO in 2015~\cite{LIGOScientific:2016aoc}, marked the beginning of a new era in observational astrophysics and cosmology. Moreover, the recent observation of the Hellings-Downs correlation in various Pulsar Timing Array (PTA) datasets~\cite{NANOGrav:2023gor,EPTA:2023fyk,Reardon:2023gzh,Xu:2023wog} has provided the first evidence of the existence of a Stochastic Gravitational Wave Background (SGWB) in the nHz frequencies range. Super Massive Black Hole (SMBH) binary systems could offer an astrophysical explanation ~\cite{1995ApJ...446..543R,Wyithe_2003,Jaffe_2003}, although they require strong deviations from standard astrophysical expectations~\cite{NANOGrav:2023hfp} and  searches for individual SMBH binaries have so far provided no positive indications~\cite{NANOGrav:2023pdq}. \nano published a study of possible new physics cosmological explanations of the data~\cite{NANOGrav:2023hvm}, including cosmic inflation, scalar-induced GWs, metastable cosmic strings, domain walls, and First Order Phase Transitions (FOPTs). Data seem to prefer a cosmological origin~\cite{NANOGrav:2023hvm} and many studies have focused on this possibility~\cite{Ellis:2023oxs, Figueroa:2023zhu, Madge:2023dxc, Wu:2023hsa, Ellis:2024xym,Bringmann:2023opz}. 
FOPTs~\cite{Gouttenoire:2023bqy, Chen:2023bms, Wang:2023bbc, Ghosh:2023aum, Addazi:2023jvg, Bringmann:2023opz, Croon:2024mde, Winkler:2024olr, Conaci:2024tlc} have emerged as one of the preferred options, fitting the data better than SMBHs due to the large amplitude and spectral shape of the signal~\cite{Winkler:2024olr} and also better than domain walls~\cite{Wu:2023hsa}.

During a FOPT, the Universe undergoes a discontinuous change in its vacuum state, driven by a scalar taking a vacuum expectation value (vev), leading to the nucleation of bubbles and the production of GWs. 
%
%
PTA results point towards a scale for the vev in the hundreds of MeVs~\cite{NANOGrav:2023hvm}, suggesting the existence of physics Beyond the Standard Model (BSM) at these energies, the so-called Dark Sectors (DS).
While for many decades the research focus has been directed towards BSM theories at high scales, recently DS have gained increasing attention~\cite{Agrawal:2021dbo, Antel:2023hkf}. They extend the SM below the EW scale and can couple to the SM via renormalizable terms, called portals: the vector portal, arising from kinetic mixing between a new dark photon and the SM gauge bosons, the neutrino one, for mixing between neutrinos and new neutral fermions, and the scalar one, in the presence of couplings between the Higgs and the dark scalar(s). Experimentally, it is established that portal couplings need to be very small, hence the name of ``dark sectors". A broad program of experimental searches is undergoing~\cite{Agrawal:2021dbo, Antel:2023hkf, CMS:2024zqs}. While major focus has been put on minimal models in which the smallest number of ingredients is added to the SM, DS models could exhibit a more complex structure, with multiple particles and interactions and large dark couplings, see e.g. the three-portal model and other examples~\cite{Ballett:2018ynz, Ballett:2019pyw, Bertuzzo:2018itn, Ballett:2019bgd, Abdullahi:2020nyr}. 

DSs are a major candidate to explain outstanding questions in cosmology and particle physics, namely the generation of neutrino masses~\cite{Bertuzzo:2018ftf, Ballett:2019cqp, Yang:2020tax, deBoer:2020yyw, Berbig:2024uwm}, baryogenesis~\cite{Haba:2010bm, Carena:2018cjh, Carena:2022qpf, Easa:2022vcw}, and dark matter~\cite{Eble:2024ivk}. 
GWs from FOPTs provide a new window on dark sectors, and in particular its dark scalar(s), making them testable even in the case in which the portal couplings are small and their phenomenological signatures suppressed. 
%
%
Studies have shown that PTA data call for a {\em supercooled} FOPT, meaning that its timescale has to be slow compared to the Hubble expansion rate. This introduces problems in the computation and in the reliability of the results. As~\cite{Athron:2023mer} has pointed out, a lack of precision in the subtleties of supercooled PT can lead to predicting FOPTs in cases where there is actually no transition at all.

In our study, we ask if, in theoretically-consistent particle physics models, in particular DS ones, it is possible to obtain the SGWB observed in PTA experiments via a PT. We take the minimal approach, considering a DS model with the fewest ingredients needed for the PT, namely a $U(1)$ gauge extension and a related scalar. In order to answer the question on its feasibility, we address the challenges posed by supercooling and go beyond previous analysis \cite{Borah:2021ocu,Ertas:2021xeh} by studying the percolation and completion temperature, using the next-to-leading order correction to the bubble wall pressure determining the correct efficiency factors. We ensure that the reheating temperature is not larger than the critical temperature. We also compute explicitly the sound speed and do not rely on the bag model approximation. Finally, 
we do not employ the $\beta$ parameter for the evaluation of the GW spectrum but we use the mean bubble separation $R_*$ which is the physically relevant quantity used in simulations.

This letter is organized as follows: firstly we introduce the model and provide an analytical understanding for the interesting region where numerically supercooling is reached. Then we discuss the various FOPT milestones the PT needs to reach until completion and the resulting GW spectrum. Finally, we present the main results and then we conclude our work.

\titleprl{Dark Scalar minimal model} We extend the Standard Model with a light DS comprised of a complex scalar, $\phi$, charged under a new $U(1)_\mathrm{D}$ gauge symmetry, and the associated dark gauge boson $Z^{\prime}_{\mu}$. The DS Lagrangian is given by 
 \begin{equation}
 \begin{split}
\mathcal{L}=&\left(D_{\mu}\phi\right)^{*}\left(D^{\mu}\phi\right)-V(\phi^*\phi) -\frac{1}{4}Z^{\prime}_{\mu\nu}Z^{\prime \mu\nu}\,,
\label{eq:lag_UV}
\end{split}
\end{equation}
%
where 
$D_{\mu}\equiv \partial_{\mu}-i\sqrt{2}g_D Z^{\prime}_{\mu}$ is the covariant derivative and $g_D$ represents the DS gauge coupling. The most general gauge-invariant scalar potential is
\begin{equation}
\begin{split}
V=&-\mu_{\phi}^2\phi^*\phi+\lambda_{\phi}\left(\phi^*\phi\right)^2\,.
\label{eq:scalar_pot_UV}
\end{split}
\end{equation}
When the dark scalar gets a vev $v_\phi=\mu_{\phi} / \sqrt{\lambda_\phi}$, the masses of the $Z^{\prime}$ and the scalar become $m_{Z'}^2 = g_D^2 v_\phi^2$ and $m_{\phi}^2 = 2\lambda_{\phi} v_\phi^2$, respectively.

The dark sector is connected to the SM via portal interactions, in particular the scalar, $\lambda_{\mathcal{H}\phi} \mathcal{H}^\dagger \mathcal{H} \phi^\dagger \phi$, and vector, $\epsilon/c_\mathrm{W} \ Z^\prime_{\mu \nu} B^{\mu \nu}$, with $B^\mu$ the SM gauge boson acting as portal and $c_\mathrm{W}$ the cosine of the Weinberg angle. 

The presence of these portal couplings is important to keep thermal equilibrium between the DS and the SM, but they are required to be small at the scales of interest by laboratory constraints~\cite{Antel:2023hkf}\footnote{Considering the relevant masses in our work, the Higgs portal interaction thermalization occurs for $\lambda_{\mathcal{H}\phi} \gtrsim 10^{-7}$~\cite{Lebedev:2021xey}, while for the dark photon $\epsilon \gtrsim 10^{-9}$~\cite{Redondo:2008ec}.}. 
While the mixed scalar coupling $\lambda_{\mathcal{H}\phi}$ could change both EW and DS symmetry breaking dynamics, its small value and the hierarchy between the EW and DS scales allows us to study the FOPT in the DS alone. Thus, in the following we study the scalar potential in Eq.~(\ref{eq:scalar_pot_UV}), 
%
adding to it the relevant one-loop corrections, namely the Coleman-Weinberg potential ($V_\mathrm{CW}$)~\cite{Coleman:1973jx} in the on-shell renormalization scheme~\cite{Anderson:1991zb, Delaunay:2007wb} and the one-loop thermal potential ($V_\mathrm{T}$)~\cite{Dolan:1973qd} including the Daisy corrections ($V_\mathrm{D}$)~\cite{Carrington:1991hz}, so that the total temperature-dependent effective potential is
%
\begin{equation}
V_{\mathrm{eff}}\left(T\right) = V + V_{\mathrm{CW}} + V_T + V_{\mathrm{daisy}}\,.
\label{eq:tota_potential}
\end{equation}
%

The Lagrangian could also contain new fermions, both charged and neutral with respect to the new gauge symmetry, hence dark fermions and heavy neutral leptons, with Yukawa couplings among themselves and with the SM leptonic doublet and the Higgs. While these terms would lead to mixing of the neutral fermions, after EW and $U(1)_D$ symmetry breaking, and could explain neutrino masses via a see-saw mechanism~\cite{Minkowski:1977sc,Mohapatra:1979ia,Yanagida:1979as,Gell-Mann:1979vob,Branco:1988ex,Kersten:2007vk,Abada:2007ux,Moffat:2017feq}, 
we neglect them in our present study focused on a minimal approach. 


\titleprl{Supercooling} In order to account for \nano measurements, we need to look for a supercooled FOPT. 
%
%
We find that regions with the highest degree of supercooling correspond to values of $g_D$ for which a barrier is present at zero temperature, given by 
\begin{equation}
g_{D}^{\mathrm{roll}} = \left\lbrace \frac{16\pi^2\lambda_{\phi}}{3}\left[1-\frac{\lambda_{\phi}}{8\pi^2}\left(5+2\log{2}\right)\right]\right\rbrace^{1/4}\,.
\label{eq:g_roll}
\end{equation}
Our numerical analysis shows that a strong and supercooled phase transition happens in the region $ g_D \sim g_D^{\rm roll}$.


\titleprl{Milestones in the FOPT}
Given the effective potential from Eq.~(\ref{eq:tota_potential}), at very high-temperatures we have symmetry restoration with only one minimum at $\varphi=0$. As the temperature decreases a second minimum appears. At the critical temperature $T_{\mathrm{crit}}$, estimated as $T_{\mathrm{crit}}^2\sim \mu^2 _{\phi} / (\lambda_{\phi}+g_D^2/2)$, the two minima become degenerate, allowing for a transition from the false to the true vacuum for $T<T_{\mathrm{crit}}$. 
At the so-called nucleation temperature $T_N$, there is on average one nucleated bubble per Hubble volume, with the Hubble expansion rate given by
%
\begin{equation}
    H(T)=\left[\frac{g_*(T)T^4}{90\pi^2M_{\mathrm{Pl}}^2}+\frac{\Delta V_{\mathrm{eff}}(T)}{3M^2_{\mathrm{Pl}}}\right]^{1/2}\,,
    \label{eq:Hubble}
\end{equation}
%
where we include the radiation contribution and the vacuum energy released in the transition. The latter is given by $\Delta V_{\mathrm{eff}}(T) = V_{\mathrm{eff}}(0,T)-V_{\mathrm{eff}}(v_{\phi,T},T)$ with $v_{\phi,T}$ the broken minimum at temperature $T$. The transition probability is controlled by the action of the bounce solution~\cite{Coleman:1973jx,Coleman:1977py,Linde:1980tt} $S_3(T)$ and is given by 
\begin{equation}
\Gamma(T)\simeq T^4\left(\frac{S_3}{2\pi T}\right)^{3/2}e^{-S_3/T}\,.
\label{eq:nuc_rate}
\end{equation}
The nucleated bubbles expand in the false vacuum until they start to collide and coalesce. The percolation temperature $T_p$ is defined when there exists a connected region of true vacuum over the entire Universe, and numerical studies~\cite{Athron:2023xlk} identify this as the moment when the remaining fraction of the Universe in the false vacuum $\mathcal{P}_f(T)$ is around 71~\%. This is the temperature at which the thermodynamic parameters that characterize the FOPT are computed. Although $T_N\sim T_p$ for fast transitions, it was shown that in the supercooling regime a nucleation temperature does not always necessarily exist~\cite{Athron:2022mmm}. Thus, we focus here solely on the percolation temperature, obtained by solving~\cite{Athron:2023xlk}
%
%
%
\begin{equation}
\begin{split}
 \int_{T_p}^{T_{\mathrm{crit}}}dT^\prime \frac{\Gamma(T^{\prime})\mathcal{V}(T_p,T^{\prime})}{\mathcal{J}(T^{\prime})H(T^{\prime})}e^{-3\int_{T_p}^{T^{\prime}}dT^{\prime\prime}/\mathcal{J}(T^{\prime\prime})}\simeq 0.31\,,
\end{split}
\label{eq:percolation_temperature}
\end{equation}
where the volume of a bubble $\mathcal{V}$ is given by
\begin{equation}
    \mathcal{V}(T_p,T) \equiv \frac{4\pi v_w^3}{3}\left(\int_{T_p}^{T}dT^{\prime}\frac{e^{\int_{T_p}^{T^{\prime}}dT^{\prime\prime}/\mathcal{J}(T^{\prime\prime})}}{\mathcal{J}(T^{\prime})H(T^{\prime})}\right)^3\,.
    \label{eq:volume_bubble}
\end{equation}
In Eq.~(\ref{eq:volume_bubble}), $v_w$ is the velocity of the bubble wall and we have used the fact that, in full generality, the relation between time and temperature is given by
\begin{equation}
\label{eq:dTdt}
\frac{dT}{dt} = - H(T) \mathcal{J}(T)\equiv - 3 H \frac{\partial V_{\mathrm{eff}}/\partial T}{\partial^2 V_{\mathrm{eff}}/\partial T^2}\Bigg|_{\varphi=0}\,.
\end{equation}
%
In the bag model, approximation usually assumed in the literature, $\mathcal{J}(T)=T$ such that the exponential factors in Eqs.~(\ref{eq:percolation_temperature}, \ref{eq:volume_bubble}) drop out. Reaching a percolation temperature is not enough, one needs to explicitly check whether the transition completes. Otherwise the PT might reach percolation but never fill completely the Universe. We ensure completion demanding that the fraction of the Universe in the false vacuum becomes as small as 1~\% for some $T_c<T_p$~\cite{Athron:2023mer} and that $\mathcal{P}_f(T)$ is still decreasing at $T=T_c$.

Right after percolation, the dark scalar decays, reheating the plasma in the regions of collision of growing bubble walls.
Assuming a fast decay rate, $\Gamma_{\varphi}>H(T_p)$,\footnote{We have checked that in more complete models, for example including new fermions, the dark scalar decays sufficiently fast and at big bang nucleosynthesis the thermal plasma contains only SM degrees of freedom. This however does not change the dynamics of the phase transition itself and thus we do not specify a particular realization. 
} such that after the transition the Universe is again radiation-dominated and a fast thermalization of the decay products, the reheating temperature, $T_{\mathrm{RH}}$, can be approximated by~\cite{Ellis:2018mja}
\begin{equation}
    T_{\mathrm{RH}}\simeq T_p\left( 1+\alpha\right)^{1/4}\,,
    \label{eq:reheating_temperature}
\end{equation}
where $\alpha$, defined in Eq.~(\ref{eq:alpha_trace_anomaly}), measures the strength of the transition. This is a good approximation for bubbles that expand as supersonic detonations\footnote{This is a good approximation for strong supercooled PTs~\cite{Leitao:2015fmj, Megevand:2016lpr, Kobakhidze:2017mru, Cai:2017tmh, Ellis:2018mja, Ellis:2019oqb, Ellis:2020nnr, Wang:2020jrd, Athron:2022mmm, Athron:2023xlk}.} in which case it is not necessary to take into account the subtleties of the reheating processes~\cite{Athron:2022mmm}. When computing the expected SGWB, the GWs are redshifted from $T_{\mathrm{RH}}$ until today~\cite{Cai:2017tmh}.


\titleprl{Gravitational waves sourced by FOPT}\label{sec:GW_params}
The thermal parameters characterizing the GW production from a FOPT are given by the reheating temperature $T_{\mathrm{RH}}$, the strength of the transition $\alpha$, and the mean bubble separation at percolation $R_*$. 

Following Refs.~\cite{Athron:2023xlk,Giese:2020rtr,Giese:2020znk}, we compute the strength of the transition by explicitly computing the speed of sound in the plasma in the broken phase, $c_{s,t}^2(T)$, and using the pseudotrace $\bar{\theta} = (\rho-3p/c_{s,t}^2)/4$, and the entalpy density $w = - T (\partial V/\partial T)$, as
\begin{equation}
    \alpha = \frac{4}{3}\frac{\bar{\theta}_f(T_p)-\bar{\theta}_t(T_p)}{w_f(T_p)}\,,
    \label{eq:alpha_trace_anomaly}
\end{equation}
where the subscripts $f\,(t)$ correspond to the value of these quantities in the false (true) minimum. This definition of $\alpha$ does not rely on the usual assumption that the bag equation of state holds, avoiding the introduction of large errors in the estimation of the kinetic energy fraction~\cite{Giese:2020rtr,Giese:2020znk}.

Finally, the mean bubble separation, which is the length-scale used in lattice simulations~\cite{Cutting:2018tjt} and from which the GW spectrum is obtained, is given by
\begin{equation}
\begin{split}
R_* &= \left[n_B(T_p)\right]^{-1/3}\\
&=\left(\int_{T_p}^{T_{\mathrm{crit}}}dT^{\prime}\frac{\Gamma(T^{\prime}) \mathcal{P}_f(T^{\prime})}{H(T^{\prime})\mathcal{\mathcal{J}(T^{\prime})}}e^{-3\int_{T_p}^{T^{\prime}}dT^{\prime\prime}/\mathcal{J}(T^{\prime\prime})}\right)^{-1/3}\,.
\end{split}
\label{eq:R_star}
\end{equation}

There are three possible sources of GWs arising from FOPT: bubble collisions dominating when the bubble walls keep accelerating through their existence, GWs sourced by sound waves mainly dominating when the walls reach a terminal velocity and thus the vacuum energy is transferred to the plasma, and finally turbulence in the plasma. 

In the following, we closely follow the GW analysis from Ref.~\cite{Ellis:2019oqb} for supercooled phase transitions, denoting with the subscript $*$ the relevant quantities computed at $T_p$. The efficiency factor for bubble collisions $\kappa_{\mathrm{col}}$, which represents the fraction of the vacuum energy that goes into the bubble walls, is given by
\begin{equation}
\begin{split}
\kappa_{\mathrm{col}}=\begin{cases}
\frac{\gamma_{\mathrm{eq}}}{\gamma_*}\left[1-\frac{\alpha_{\infty}}{\alpha}\left(\frac{\gamma_{\mathrm{eq}}}{\gamma_*}\right)^2\right]\,, & \gamma_*>\gamma_{\mathrm{eq}}\\
1-\frac{\alpha_{\infty}}{\alpha}\,, & \gamma_*\leq \gamma_{\mathrm{eq}}
\end{cases}\,.
\end{split}
\label{eq:kappa_col}
\end{equation}
It takes into account the friction terms that can drive the bubble wall to a terminal velocity, with a Lorentz factor, $\gamma_{\mathrm{eq}}$~\cite{Bodeker:2009qy, Bodeker:2017cim, Gouttenoire:2021kjv}. In Eq.~(\ref{eq:kappa_col}), $\gamma_{\mathrm{eq}}$ and $\alpha_{\infty}$ are given by
\begin{equation}
\begin{split}
    \gamma_{\mathrm{eq}}=\frac{\alpha-\alpha_{\infty}}{\alpha_{\mathrm{eq}}}\,,\;\mathrm{and}\; \alpha_{\infty} \equiv \frac{1}{18}\frac{\Delta m^2 T_p^2}{w_f(T_p)}\,,
\end{split}
\end{equation}
%
%
with $\alpha_{\mathrm{eq}}\equiv \frac{4g^2\Delta m_V T_p^3}{3w_f(T_p)}$ and $\Delta m^2\equiv \sum_i n_i(m_{i,t}^2-m_{i,f}^2)$ summed over the bosonic degrees of freedom\footnote{Note that we are not specifying any particular fermionic degrees of freedom, which would contribute similarly to bosons but with an extra 1/2 factor.} and $g^2 \Delta m_V\equiv 3g_X^2 (m_{X,t}-m_{X,f})$ only running over the gauge bosons gaining a mass in the transition. The quantity $\alpha_{\infty}$ represents the weakest transition that can take place overcoming the leading-order friction term, and when $\alpha>\alpha_{\infty}$, we are in the so-called runaway regime such that $v_w\rightarrow 1$. Instead, $\alpha_{\mathrm{eq}}$ is related to the NLO friction term that drives the expanding bubbles to a terminal velocity~\cite{Bodeker:2017cim,Gouttenoire:2021kjv}.  $\gamma_{\mathrm{eq}}$ is compared to the Lorentz factor the bubble wall would reach if the NLO pressure term was not present
%
%
%
\begin{equation}
    \gamma_* \equiv \frac{2}{3}\frac{R_*}{R_0}\,,
\end{equation}
with $R_0$ 
as defined in Ref.~\cite{Ellis:2019oqb}. Thus, if $\gamma_*\leq \gamma_{\mathrm{eq}}$, all the vacuum energy goes intro accelerating the bubble wall. When $\gamma_*> \gamma_{\mathrm{eq}}$, the leftover energy after the wall has reached its terminal velocity is instead damped in the surrounding plasma, generating GWs from sound waves and turbulence. We find that the main contribution to the GW arises from sound waves, whose power spectrum can be written as~\cite{Caprini:2015zlo}
\begin{equation}
    \begin{split}
        \Omega_{\mathrm{sw},*}&=0.38 (H_* R_*)(H_*\tau_{\mathrm{sw}})\left(\frac{\kappa_{\mathrm{sw}}\alpha}{1+\alpha}\right)^2\left(\frac{f}{f_{\mathrm{sw}}}\right)^3\\
        &\times \left[1+\frac{3}{4}\left(\frac{f}{f_{\mathrm{sw}}}\right)^2\right]^{-7/2}
    \end{split}
\end{equation}
with peak frequency $f_{\mathrm{sw}}=3.4/\left[\left(v_w-c_s\right)R_*\right]$. The efficiency coefficient for sound waves is approximately given by~\cite{Espinosa:2010hh,Caprini:2015zlo}:
\begin{equation}
    \kappa_{\mathrm{sw}}=\frac{\alpha_{\mathrm{eff}}}{\alpha}\frac{\alpha_{\mathrm{eff}}}{0.73+0.083\sqrt{\alpha_{\mathrm{eff}}}+\alpha_{\mathrm{eff}}}
\end{equation}
with $\alpha_{\mathrm{eff}}\equiv \alpha\left(1-\kappa_{\mathrm{col}}\right)$. Finally, $\tau_{\mathrm{sw}}$ is the length of the sound wave period and we follow the prescription presented in Ref.~\cite{Ellis:2019oqb}. If the sound wave period is significantly shorter than a Hubble time, then a sizable fraction of the vacuum energy goes into turbulence in the plasma.



\titleprl{Numerical analysis}
The numerical computation of the thermal parameters used as inputs in the prediction of the gravitational wave spectra requires a dedicated analysis, given that we are interested in supercooled FOPTs. We will point out the main improvements we considered over previous studies, essential to establish the reliability of the FOPT explanation of \nano data. Throughout all our discussion, we study solutions in the runaway regime ($\alpha>\alpha_{\infty}$).

We implemented the potential using the model class provided by \texttt{CosmoTransitions}~\cite{Wainwright:2011kj} and calculated the action using the tunneling potential method in~\cite{Espinosa:2018szu, Espinosa:2018hue}. 
One important feature of our analysis is that it does not rely on the bag model, which considers regions or ``bags'' of constant energy per volume~\cite{Chodos:1974je} such that the potential scales like $V_{\mathrm{eff}}(T)\sim aT^4+b$, with $a$ and $b$ constant. As can be seen from Eq.~(\ref{eq:dTdt}), this affects directly the relation between time and temperature which is used throughout the computation of every thermal parameter describing the FOPT.
Additionally, as already explained earlier, when using the bag model relations between enthalpy and energy densities, $\rho=4w/3$, the kinetic energy of the plasma can be overestimated~\cite{Giese:2020rtr, Giese:2020znk, Wang:2023bbc}. Instead, we use the pseudotrace and enthalpy densities in the computation of $\alpha$, as defined in Eq. (\ref{eq:alpha_trace_anomaly}). Additionally, we directly compute the speed of sound in the plasma $c_s^2 (T) = \partial_T V / (T \partial_T^2 V)$ at a given temperature, instead of assuming a constant $c_s=1/\sqrt{3}$ and finding deviations from this value as large as $20\%$ in some cases.


Finally, we compute explicitly the mean bubble separation at percolation $R_*$, given that lattice simulations providing with the GW spectrum are performed in terms of this length scale~\cite{Hindmarsh:2015qta,Cutting:2018tjt}. This circumvents the introduction of further assumptions on the behavior of the nucleation rate which are found when trying to relate $R_*$ with the commonly used inverse time duration of the transition $\beta_*$.

\begin{figure}
    \includegraphics[width=0.49\textwidth]{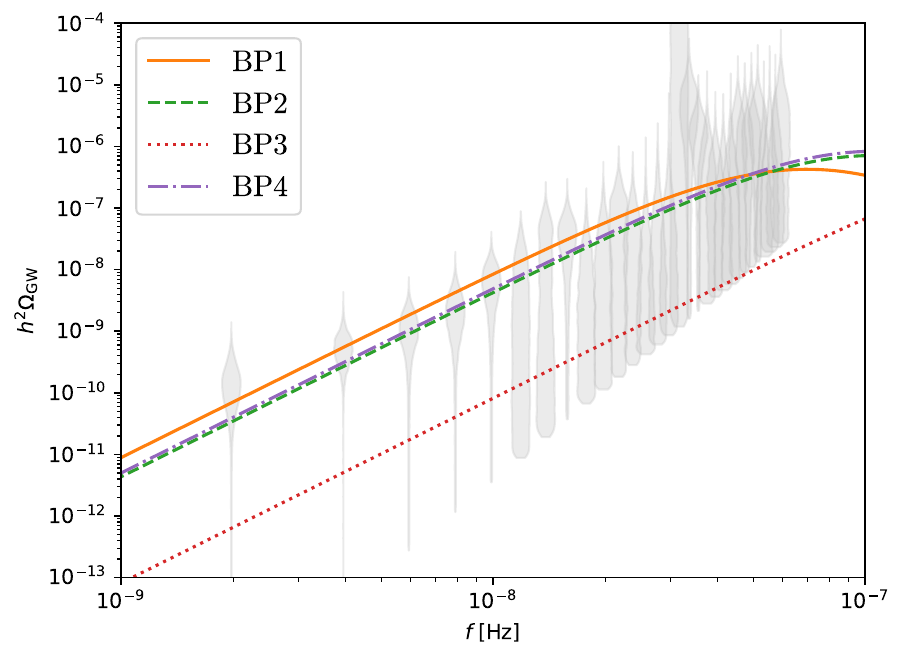}
    \caption{Gravitational wave spectrum for the reference points in Table~\ref{tab:benchmarks}. The gray shaded regions correspond to NANOGrav's violin plot. 
    \label{fig:spectra_BP1}}
\end{figure}

\titleprl{Benchmark points}
In the following we present four benchmark points (BP) in Table~\ref{tab:benchmarks} that could account for the \nano 15-year data set~\cite{NANOGrav:2023hvm}, showing indeed that it can be explained with a FOPT. As previously found~\cite{NANOGrav:2023hvm}, we also see that the vev is required to be in the $\sim 500-1000$~ MeV range. For BP1-3 we fix the couplings $\lambda_{\phi} = 0.006$ and $g_D \simeq g_D^{\mathrm{roll}}= 0.75$ studying how the GW spectrum changes varying the vev, while BP4 has the same vev as BP2 but different couplings. We leave a scan of the full parameter space for future studies. The resulting GW spectra are shown in Fig.~\ref{fig:spectra_BP1}, together with the violins from \nano 15-year data.
\begin{table}
\begin{tabular}{|c|c|c|c|c|c|c|c|c|}
\hline
 & $v_{\phi}$ & $\lambda_{\phi}$ &  $g_D$ & $T_p$ & $H_*R_*$ & $\alpha$ & $m_\varphi$ & $m_{Z^\prime}$ \\
 & [GeV] &  &  & [MeV] &  &  & [GeV] & [GeV] \\
 \hline
BP1 & 0.5  & 0.006 & 0.75 & 12.37 & 0.489 & 361.5 & 0.055  & 0.375  \\
\hline
BP2 & 1  & 0.006 & 0.75 & 23.37 & 0.651 & 451.8 & 0.110 & 0.750  \\
\hline
BP3 & 3  & 0.006 & 0.75 & 70.29 & 0.644 & 367.6 & 0.329 & 2.25 \\
\hline
BP4 & 1  & 0.010 & 0.85 & 38.39 & 0.723 & 94.92 & 0.141 & 0.854  \\
\hline
\end{tabular}
   
\caption{Benchmark points reproducing the PTA results from \nano with FOPTs with $g_D\simeq g_D^{\mathrm{roll}}$. The first three columns correspond to the physics parameter of the dark sector scenario we consider, while the next three columns correspond to the percolation temperature, the mean bubble separation at percolation and the transition strength. Finally we present the masses of the dark scalar and the dark photon.
 \label{tab:benchmarks}}
\end{table}

As stressed in previous sections, particular care needs to be taken in order to reproduce the \nano results making sure that the PT completes. Indeed, if the completion condition we impose following Ref.~\cite{Athron:2023mer} was lifted, a much larger degree of supercooling would be found. However, such thermal histories with a long period of exponential expansion make the FOPT unsuccessful as the Universe would remain in the false vacuum. Given that the completion of the FOPT limits the amount of supercooling we can have, we find that the main contribution to the GW spectrum comes from sound waves. This is due to the NLO pressure contribution from gauge-boson emission driving the wall to a terminal velocity, suppressing the contribution from bubble collisions. Bubble collisions are only the dominant GW source for very large values of $\alpha\gg 10^4$~\cite{Ellis:2019oqb}, solutions for which we find that the FOPT does not complete, in line with Ref.~\cite{Athron:2023mer}.

\begin{figure}
    \includegraphics[width=0.49\textwidth]{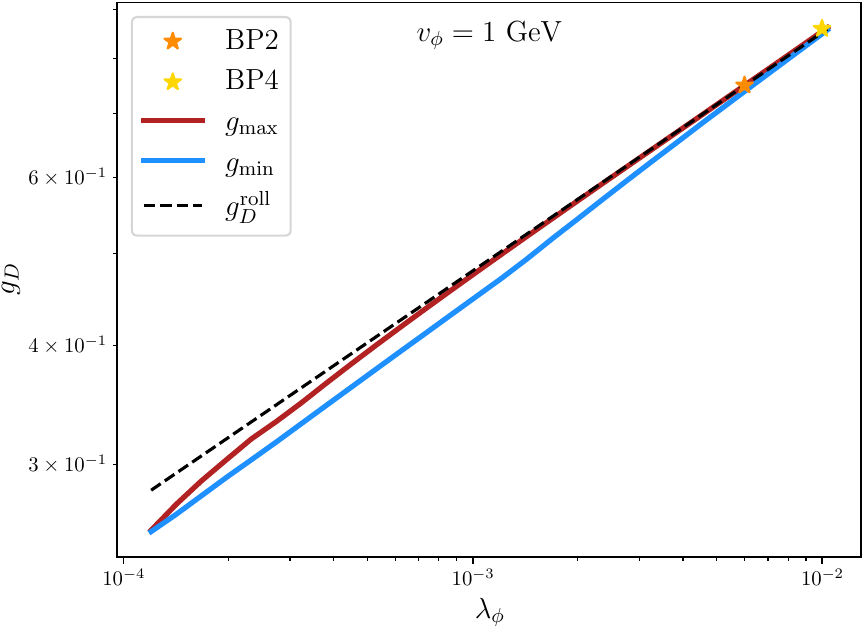}
    \includegraphics[width=0.49\textwidth]{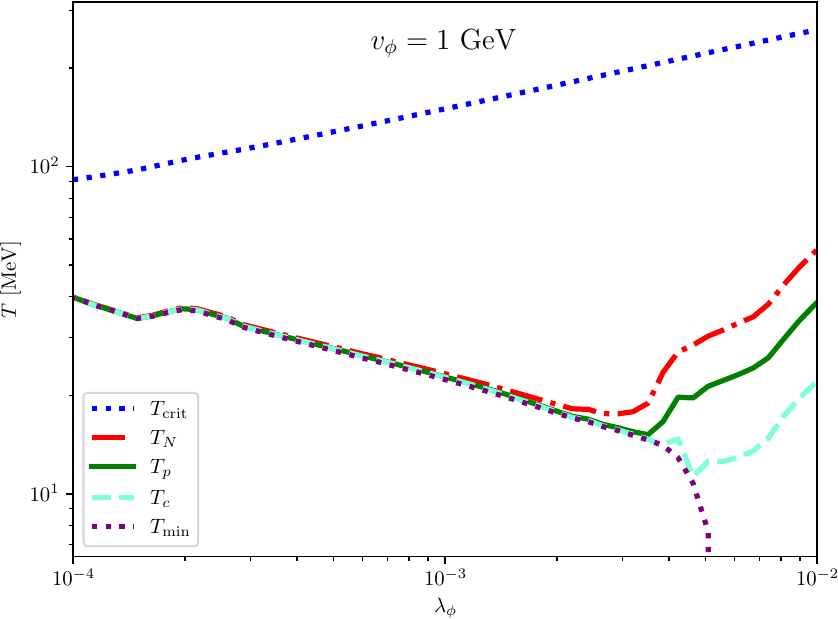}
    \caption{
    Upper panel: Maximum (minimum) value $g_\mathrm{max}$ ($g_\mathrm{min}$) of the gauge coupling for which a FOPT completes as a function of the scalar quartic coupling $\lambda_{\phi}$, shown in dark red (blue). The dashed line corresponds to the values for which a potential barrier is present at zero temperature (see Eq.~(\ref{eq:g_roll})).
    Lower panel: Values of the different relevant temperatures as a function of $\lambda_{\phi}$ for $g_D=g_D^{\mathrm{max}}$. In both panels the vev is fixed to that of BP2 and BP4 in Table~\ref{tab:benchmarks}. \label{fig:temperatures_and_groll}}
\end{figure}
In general, we find that successfully accounting for the SGWB observation corresponds to the gauge coupling values $g_D\sim g_D^{\mathrm{roll}}$, which is expected given that for this value of $g_D$ a barrier separating the minima at zero temperature allows for a higher degree of supercooling. This is shown in Fig.~\ref{fig:temperatures_and_groll} as well. In the upper panel, we show in dark red (blue) the largest (smallest) value of the gauge coupling $g_\mathrm{max}$ ($g_\mathrm{min}$), computed numerically, for which the FOPT completes in the runaway regime as a function of $\lambda_\phi$ and by fixing the vev to that of BP2. The black dashed line corresponds to the values of the gauge couplings given by Eq.~(\ref{eq:g_roll}). As can be seen in combination with the lower panel, in which we show different relevant temperatures in the FOPT as a function of $\lambda_{\phi}$ and for $g_D=g_D^{\mathrm{max}}$, the highest degree of supercooling appears when $g_D\sim g_D^{\mathrm{roll}}$. For this value, we no longer find a minimum temperature at which the FOPT can take place ($T_{\mathrm{min}}$), as a consequence of the potential barrier arising at zero temperature. Although this can be regarded as fine-tuned, it might point toward particular scenarios with classical scale invariance which enjoy similar features~\cite{Salvio:2023blb, Salvio:2023ynn, Goncalves:2025uwh}. 

We highlight that the condition for supercooling, $g_D \simeq g_D^\mathrm{roll}$, leads to a significant hierarchy between the mass of the scalar which is much lighter than the dark gauge boson, see Table~\ref{tab:benchmarks}, with potential key consequences for their phenomenology and related searches.

\titleprl{Conclusions}
We have studied the possible explanation of the \nano 15-year data data via supercooled FOPT, assuming a minimal approach.  In particular, we showed that a 100 MeV$-$GeV dark sector comprised of a complex scalar charged under a $U(1)_D$ gauge symmetry, can successfully do so. 

By means of a rigorous approach that takes care of the supercooling subtleties of the FOPT, going beyond commonly-made approximations, we find that the main contribution to the GW spectrum arises from sound waves, given that the friction arising from soft gauge boson emission drives the expansion of the walls to a terminal velocity, suppressing the contribution from bubble collisions. This is because the amount of supercooling is limited by the condition that the FOPT completes, as noted in Ref.~\cite{Athron:2023mer}.  

In order to reproduce the observed SGWB, we find a tight relation between the size of the gauge coupling and that of the quartic coupling of the light scalar. The gauge coupling values required are around $g_D^{\mathrm{roll}}$ for which a potential barrier between the minima at zero temperature is generated.  Correspondingly, the quartic coupling needs to be significantly smaller. This results in a hierarchical spectrum for the dark bosons, with a much heavier gauge boson with respect to the dark scalar that 
typically has mass in the $\mathcal{O}\left(10-1000\right)$~MeV range. 
This could lead to important phenomenological consequences once the portals to the SM are taken into account and could guide future laboratory tests of these models. Non-minimal realizations of this model could include fermions, both mixing to neutrinos as as DM, that would enrich even more the dark phenomenology.


\begin{acknowledgments}
\titleprl{Acknowledgments}
We are deeply grateful to Yann Gouttenoire, Andrea Mitridate and Tanner Trickle for their help in reproducing the \nano violin plots in Fig.~\ref{fig:spectra_BP1}. M.~L. and F.~C. thank Carlo Tasillo for helpful and interesting discussions. F.C. thanks Giulio Barni for fruitful conversations. M.~L. thanks Fermilab for hosting him during the development of this work.
The research of F.~C., S.~P. and S.~R.~A. has received funding / support from the European Union’s Horizon 2020 research and innovation programme under the Marie Skłodowska-Curie grant agreement No 860881-HIDDeN. F.~C. acknowledges support from the FORTE project CZ.02.01.01/00/22\_008/0004632 co-funded by the EU  and the Ministry of Education, Youth and Sports of the Czech Republic. J.~H.~Z. is supported by the National Science Centre, Poland (research grant No. 2021/42/E/ST2/00031). M.~L. is funded by the European Union under the Horizon Europe's Marie Sklodowska-Curie project 101068791 — NuBridge.
\end{acknowledgments}

\begin{acknowledgments}
\titleprl{Note added}
While this paper was being written, a similar study was released in Ref.~\cite{Goncalves:2025uwh}. In particular, the authors find the possibility to explain \nano results with a classically scale invariant scalar sector which they relate to $B-L$. Here we do not assume scale invariance in the dark sector and, more importantly, we avoid the use of the bag model assumptions to compute the relevant thermal parameters for the FOPT. Furthermore, we directly compute the mean bubble separation $R_*$, instead of the inverse time duration $\beta$, which relies on further assumptions on the shape of the bubble nucleation.
\end{acknowledgments}

\bibliographystyle{apsrev4-1}
\bibliography{lib}{}


\newpage
\onecolumngrid
\setlength{\parindent}{15pt}
\setlength{\parskip}{1em}
\newpage
\hypertarget{Supp_Mat}{}

\begin{center}
	\textbf{\large Supercooled Dark Scalar Phase Transitions explanation of NANOGrav data} \\
    \vspace{0.05in}
	{ \small \bf (Supplemental Material)}\\ 
	\vspace{0.05in}
    {Francesco Costa,$^1$ Jaime Hoefken Zink,$^2$ Michele Lucente,$^{3,4}$ Silvia Pascoli,$^{3,4}$ and Salvador Rosauro-Alcaraz$^{4}$}
\end{center}
\vspace{-.5em}
\centerline{{\it  $^{1}$ Institute of Particle and Nuclear Physics, Faculty of Mathematics and Physics,}}
    \centerline{{\it  V Hole\v{s}ovi\v{c}k\'ach 2, 180 00 Praha 8, Czech Republic;}}
\centerline{{\it $^{2}$ National Centre for Nuclear Research, Pasteura 7, Warsaw, PL-02-093, Poland; }}
\centerline{{\it $^{3}$ Dipartimento di Fisica e Astronomia, Universit\`a di Bologna, via Irnerio 46, 40126 Bologna, Italy; and}}
\centerline{{\it $^{4}$ INFN, Sezione di Bologna, viale Berti Pichat 6/2, 40127 Bologna, Italy }}
\vspace{0.05in}
\setcounter{page}{1}

\section{One-loop effective potential}
\label{app:effective potential}
In order to study the phase transition dynamics and the generation of GW, we will use the one-loop effective potential (written in terms of the background field $\phi\rightarrow \varphi/\sqrt{2}$) which can be decomposed as
\begin{equation}
V_{\mathrm{eff}}\left(\varphi,T\right) = V_{\mathrm{tree}}+V_{\mathrm{CW}}+V_T+V_{\mathrm{daisy}}\,,
\label{eq:total_potential}
\end{equation}
where the first two terms correspond, respectively,  to the tree-level potential 
\begin{equation}
V_{\mathrm{tree}}\left(\varphi\right) = -\frac{1}{2} \mu_{\phi}^2 \varphi^2 + \frac{\lambda_{\phi}}{4} \varphi^4\,,
\label{eq:tree_level_pot}
\end{equation}
and the Coleman-Weinberg (CW) potential given by
\begin{equation}
\begin{split}
V_{\mathrm{CW}}\left(\varphi\right)=&\sum_i\frac{n_i}{64\pi^2}\Bigg[m_i^4\left(\varphi\right)\left(\log\frac{m_i^2\left(\varphi\right)}{m_i^2\left(v_{\phi}\right)}-\frac{3}{2}\right)
+2m_i^2\left(\varphi\right)m_i^2\left(v_\phi\right)\Bigg]\,,
\end{split}
\label{eq:CW_pot}
\end{equation}
where 
$n_i$ are the number of degrees of freedom for each species. Note that the one-loop potential from Eq.~(\ref{eq:CW_pot}) guarantees that loop effects do not change neither the vev nor the scalar mass with respect to the tree-level values at zero temperature. The temperature dependence is encoded in the terms $V_T$ and $V_{\mathrm{daisy}}$. The first is given by
\begin{equation}
\begin{split}
V_T\left(\varphi,T\right)=&\sum_{i=\mathrm{bosons}}\frac{n_i T^4}{2\pi^2}J_B\left(\frac{m_i^2\left(\varphi\right)}{T^2}\right)
+\sum_{i=\mathrm{fermions}}\frac{n_i T^4}{2\pi^2}J_F\left(\frac{m_i^2\left(\varphi\right)}{T^2}\right)\,,
\end{split}
\label{eq:1loopT_pot}
\end{equation}
where the function $J_{B(F)}$ is given by
\begin{equation}
J_{B(F)}(x)\equiv \pm \int_0^{\infty}dy y^2\log{\left(1\mp e^{-\sqrt{x^2+y^2}}\right)}\,,
\end{equation}
for bosons and fermions, respectively. The last part of the one-loop potential is the contribution from Daisy resummation, $V_{\mathrm{daisy}}$, which using the Arnold-Espinosa prescription~\cite{Arnold:1992rz} is found to be
\begin{equation}
V_{\mathrm{daisy}}\left(\varphi,T\right)=\sum_{i=\mathrm{bosons}}\frac{T\tilde{n}_i}{12\pi}\left[m_i^3\left(\varphi\right)-\left(m_i^2\left(\varphi\right)+\Pi_i\left(T\right)\right)^{3/2}\right]\,,
\label{eq:daisy_pot}
\end{equation}
where $\Pi_i\left(T\right)$ is the thermal mass in the high-temperature limit, and $\tilde{n}_i$ is the number of degrees of freedom that develop a thermal mass. Let us underline that only the longitudinal mode of the gauge bosons gets a thermal mass, and thus $\tilde{n}_i=1$, while in general for a massive vector boson one has $n_i=3$. We find the following contributions to the boson thermal masses:
\begin{equation}
\begin{split}
\Pi_\varphi (T) =&\left(\lambda_{\phi}+\frac{g_X^2}{2}\right)\frac{T^2}{2}\,,\;\mathrm{and}\;
\Pi_{Z^{\prime}} (T) =\frac{g_D^2}{3}T^2\,,
\end{split}
\label{eq:thermal_masses}
\end{equation}
while the corresponding thermal mass for the Goldstone boson matches $\Pi_{\varphi}(T)$. Given that $\Pi_{i} (T)$ is computed in the high-temperature regime, we include a cut-off on its size given by~\cite{Baldes:2018emh}:
\begin{equation}
\label{eq:Daisy_cutoff}
\begin{split}
\Pi_{i} (T)  \to \Pi_{i} (T) x^2 K_2 (x) / 2
\end{split}
\end{equation}
where $x \equiv m_i /T $ and $K_2$ is the modified Bessel function of the second kind.

We estimate that the running of the gauge coupling, not included in our analysis, induces percent-level changes, while the gauge boson mass may vary by a factor of a few~\cite{Zhu:2025pht,Kierkla:2022odc}. Although important for detailed parameter space studies, this effect does not impact our main focus: the generation of the PTA GW signal and the obtained relation between gauge and scalar masses.

\section{GW from collisions and turbulence}
The contribution to the GW signal from bubble collisions is given by~\cite{Cutting:2018tjt,Ellis:2019oqb}
\begin{equation}
\begin{split}
\Omega_{\mathrm{col},*}&=0.024(H_*R_*)^2\left(\frac{\kappa_{\mathrm{col}}\alpha}{1+\alpha}\right)^2\left(\frac{f}{f_{\mathrm{col}}}\right)^3 
 \left[1+2\left(\frac{f}{f_{\mathrm{col}}}\right)^{2.07}\right]^{-2.18}\,,
\end{split}
\label{eq:omega_col}
\end{equation}
with $f_{\mathrm{col}}=0.51/R_*$. When the bubble walls reach a terminal velocity, as found in our study, the main contributions to the GW spectrum are sound waves and turbulance. If the sound wave period is significantly shorter than a Hubble time, then a sizable fraction of the vacuum energy goes into turbulence in the plasma, which sources a GW spectrum given by~\cite{Caprini:2009yp}
\begin{equation}
\begin{split}
    \Omega_{\mathrm{turb},*}&=6.8(H_*R_*)\frac{(1-H_*\tau_{\mathrm{sw}})}{1+8\pi f/H_*}\left(\frac{\kappa_{\mathrm{sw}}\alpha}{1+\alpha}\right)^{3/2}
     \left(\frac{f}{f_{\mathrm{turb}}}\right)^3\left[1+\left(\frac{f}{f_{\mathrm{turb}}}\right)\right]^{-11/3}
\end{split}
\end{equation}
and whose peak frequency corresponds to $f_{\mathrm{turb}}=3.9/\left[\left(v_w-c_s\right)R_*\right]$. The total GW spectrum is obtained as $\Omega_{\mathrm{GW},*}=\Omega_{\mathrm{col},*}+\Omega_{\mathrm{sw},*}+\Omega_{\mathrm{turb},*}$, although $\Omega_{\mathrm{sw},*}$ dominates the signal explaining NANOGrav 15-year data.

\section{FOPT Completion}
\begin{figure}[h]
    \includegraphics[width=0.49\textwidth]{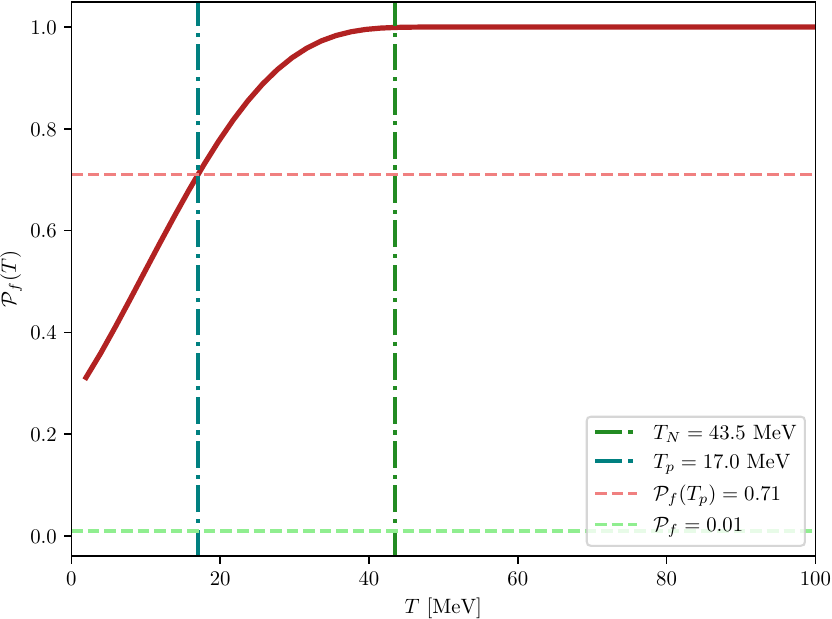}
    \includegraphics[width=0.49\textwidth]{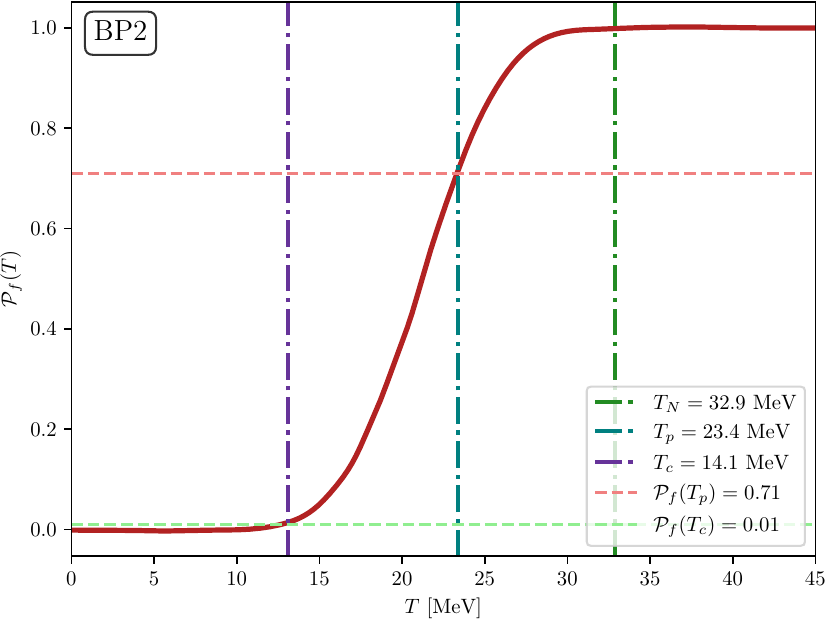}
    \caption{Evolution of the false vacuum fraction of the Universe $\mathcal{P}_f$ as a function of the temperature, shown in dark red. The dark green, teal and purple dash-dotted vertical lines correspond to the nucleation, percolation and completion temperatures, respectively. The horizontal coral and light-green dashed lines correspond to the values of $\mathcal{P}_f(T)$ that define the percolation and completion temperatures, respectively. The left panel corresponds to a point in which the FOPT does not complete, while the right one is BP2 in Table~\ref{tab:benchmarks}. \label{fig:Completions_stories}}
\end{figure}

We discussed in the main text the importance to check that the false vacuum fraction decreases making possible the completion of the FOPT. In Figure~\ref{fig:Completions_stories} we present the false vacuum fraction evolution with respect to the temperature as a dark red line, for one point which has strong supercooling but does not complete (left panel) and BP2 in Table~\ref{tab:benchmarks}. The vertical lines are the relevant temperature milestones of the PT as specified in the legend. We immediately see that for the left hand side plot the PT never completes since $\mathcal{P}_f(T)$ decreases too slowly and never reaches values close to zero. While on the right the PT reaches completion, with the false vacuum fraction reaching the values of $\mathcal{P}_f(T) \sim 0.01$ in a finite time and having a negative derivative at that temperature.

\end{document}